\documentclass[12pt]{iopart}
\usepackage{iopams}
\newcommand{\Eins}
           {\;\smash{\raisebox{-0.5ex}{$\!\!\stackrel{\!\mbox{1}
            \hspace{-0.4ex}\rule[0.0ex]{0.06ex}{1.60ex}}{ }$}}}
\newtheorem{lemma}{Lemma}
\newtheorem{defi}{Definition}
\newtheorem{prop}{Proposition}

\begin{document}

\title[Isospectral spin systems]{Continuous families of isospectral Heisenberg spin systems and the limits of
inference from measurements}

\author{Heinz-J\"urgen Schmidt\dag
\footnote[3]{To
whom correspondence should be addressed (hschmidt@uos.de)}
and Marshall Luban\ddag}

\address{\dag\ Universit\"at Osnabr\"uck, Fachbereich Physik,
Barbarastr. 7, 49069 Osnabr\"uck, Germany}

\address{\ddag\ Ames Laboratory and Department of Physics and Astronomy, Iowa State University\\
Ames, Iowa 50011, USA}

\begin{abstract}
We investigate classes of quantum Heisenberg spin systems which have different
coupling constants but the same energy spectrum and hence the same thermodynamical properties.
To this end we define various types of isospectrality and establish conditions for
their occurence. The triangle and the
tetrahedron whose vertices are occupied by spins $\frac{1}{2}$ are investigated in some detail.
The problem is also of
practical interest since isospectrality presents an obstacle to the experimental determination
of the coupling constants of small interacting spin systems such as
magnetic molecules.

\end{abstract}



\maketitle

\section{Introduction}

The measurement of the temperature-dependent magnetic susceptibility,
$\chi(T)$, provides a standard essential diagnostic method for establishing
the magnetic properties of a system. A careful comparison between
measured data and the predictions for $\chi(T)$ as derived from a model
Hamiltonian is routinely performed with the goal of establishing
numerical values of model parameters, for example, the exchange
constant(s) of the Heisenberg model of interacting isolated spins.
The success of this technique is so firmly established that it is
taken for granted that there is a one-to-one correspondence between a
given form of $\chi(T)$ and the numerical values of the model parameters. Most
certainly it is unnatural to contemplate that one might be able to
continuously vary the parameters of a model Hamiltonian and yet
generate a single, invariant form for $\chi(T)$ and similarly for other
thermodynamic quantities. Yet surprisingly,
there are a number of
exceptional systems, where there is a continuous infinity-to-one
correspondence between model Hamiltonians and measurable
thermodynamic quantities.
One of these exceptional cases was
recently encountered \cite{Jun},\cite{Lub} in the course of attempting to determine the
exchange constants of a simulating Heisenberg model from experimental
susceptibility data for a specific synthetic magnetic compound \cite{MMBSB}. \\

We shall refer to systems of a continuous
family
having the same eigenvalue spectrum as being ``isospectral".
This notion is chosen in analogy to the use of
``isospectrality" in other areas of
physics, e.~g.~the occurrence of supersymmetric pairs of Hamiltonians
(see, e.~g.~\cite{CKS}, 7.1) or the problem
of bounded domains with isospectral Laplacians
(``Can one hear the shape of a drum ?" \cite{Kac}).
The subject of isospectral spin systems
is not completely novel but has been discussed in the
literature only on a few  occasions, e.~g.~\cite{G1}, \cite{G2}.
However, there is, to our best knowledge, no systematic account of this phenomenon,
the first steps of which will be presented in this paper.\\

In addition to providing a general approach to isospectrality
we analyze in depth two cases of distinct Heisenberg systems where
continuous variation of the exchange constants gives rise to one and
the same set of temperature-dependent thermodynamic quantities. The
operational conclusion for an experimentalist is quite sobering, in
that for these specific systems measured data alone cannot fix the exchange
constants. Comparison between theory and experiment can only place a
weak constraint on a continuous family of equally acceptable choices
of parameters. Although we provide some helpful insights, it is very
difficult to formulate the general set of conditions to be met so as
to achieve such exceptional model systems. It is reasonable to
expect, that if one and the same temperature
dependent thermodynamic quantity is generated by a continuous family
of Hamiltonians then necessarily all members of that family share the
very same eigenvalue spectrum. This is indeed the case as it is
proven in section 3 for two particular thermodynamic functions.\\

In short, our goal in the present work is to provide a first systematic
study of Heisenberg isospectral spin systems.
The remarkable advances \cite{Ga},\cite{M} in synthesis magnetochemistry,
of incorporating significant numbers of interacting paramagnetic
centers within individual molecules, may provide the
impetus for wider studies that will yield a more comprehensive set of
conditions for the occurence of isospectral spin systems.\\

The layout of this paper is as follows:\\
In section 2 we introduce our notation and the basic concepts of
``isospectrality", ``complete isospectrality", and ``covariant isospectrality"
for spin systems with Heisenberg Hamiltonians. Families of isospectral systems are algebraic
varieties in the space $\cal J$ of coupling constants. Covariant isospectrality is
implemented by a unitary representation of some Lie group,
which simplifies the calculations considerably.
Unfortunately this is a rare case, as we will see.
Complete isospectrality means that all eigenvalues of two systems with the same
magnetic quantum number are in 1:1 correspondence and equal.
We do not know whether this is a strictly stronger property
than plain isospectrality, except for the case of a trivial counter-example.
However, we need this apparently
stronger concept to derive the conclusion that
completely isospectral systems share the same
magnetic susceptibility function. This is done in section 3 where also the inverse
problem is settled as well as the analogous question for the specific heat function.
The result
in short is the following: Plain isospectrality is equivalent to possessing the same
specific heat function and necessary for possessing the same magnetic
susceptibility function. Complete isospectrality is sufficient for possessing the same magnetic
susceptibility function.
In section 4 we identify the isospectral invariants which are linear or quadratic in the coupling
constants. This is crucial for section 5 where we show that the
triangle with spin $s=\frac{1}{2}$ is both of completely and
covariantly isospectral type but that for $s>\frac{1}{2}$ isospectrality breaks down.
On the other side, if the number $N$ of spin sites exceeds $3$, covariant isospectrality
is no longer possible. This is proved in section 6 with the aid of MATHEMATICA$^{\circledR}$ 4.0 and some
trace formulae which are explained in Appendix A. The tetrahedron ($N=4$) with $s=\frac{1}{2}$
nevertheless possesses completely isospectral families of dimension one and two,
as is shown in section 7 and Appendix B.  In section 8 we provide some
heuristic arguments in order to explain the findings of the previous sections. We expect
that isospectrality only occurs if the number of all possible bonds exceeds the number
of independent eigenvalues and show that this never happens except for $s=\frac{1}{2}$ and
$N=3,4,5$.
A table summarizing our results and conjectures on the occurence of isospectrality for
different $N$ and $s$ and concluding remarks are provided in section 9.

\section{Notations and definitions}

We consider spin systems with $N$ spin sites, spin quantum number $s$
and isotropic Heisenberg coupling
between all sites $x$ and $y$ with coupling constants $J_{xy}$.
For sake of compact notation we will write the ${N\choose 2}$ coupling constants
$J_{xy}$ as the components of a vector $\vec{J}\in{\cal J}={\Bbb R}^{N\choose 2}$.
Thus a specific point of ${\cal J}$ uniquely specifies the strength of the interactions between
all pairs of spins and will be sometimes called a ``system".\\

Let $S_{x}^{(i)}, (i=1,2,3),$ denote the three components of the spin observable
$\bi{S}_x$ at site $x$ and, as usual,
\begin{equation}\label{1}
\bi{S}=\sum_x \bi{S}_x,\quad S^{(i)}=\sum_x S^{(i)}_x,\quad S^\pm=S^{(1)}\pm i S^{(2)},
\end{equation}
denote the total spin vector and its various components. All linear operators
occuring in this context will be identified with the corresponding $dim\times dim$-matrices,
$dim=(2s+1)^N$ being the dimension of the total Hilbert space of the spin system,
w.~r.~t.~the fixed basis consisting of tensor products of eigenvectors of $S_x^{(3)}$. The
Hamilton operator can then be written as
\begin{equation}\label{2}
H_0 = \bi{J}\cdot\bi{H}=\sum_{x<y}J_{xy} H_{xy},
\end{equation}
where
\begin{equation}\label{3}
H_{xy}=\bi{S}_x\cdot\bi{S}_y = \sum_{i=1}^3 S_x^{(i)}S_y^{(i)}.
\end{equation}
Here $\bi{H}$ is an ${N\choose 2}$-dimensional vector the components of which are
$dim\times dim$-dimensional matrices $H_{xy}$.\\

If the spin system is coupled to a constant external magnetic field $\cal H$, the
total Hamilton operator will be
\begin{equation}\label{4}
H(h)=H_0 - h S^{(3)},
\end{equation}
where $h\equiv  g\mu_B {\cal H}$ contains the common combination of the gyromagnetic ratio $g$
and the Bohr magneton $\mu_B$. As usual, the partition function,
which yields all the standard thermodynamic functions, is defined by
\begin{equation}\label{5}
{\cal Z}(\beta,h)\equiv  \Tr (\exp({-\beta H(h))}).
\end{equation}
In particular, one obtains from ${\cal Z}$ the specific heat function
\begin{equation}\label{5a}
c(\beta)=\beta^2\frac{\partial^2}{\partial\beta^2}\ln {\cal Z}(\beta,0),
\end{equation}
and the magnetic zero field susceptibility function
\begin{equation}\label{5b}
\chi(\beta)=\frac{1}{\beta}\frac{\partial^2}{\partial h^2}\ln {\cal Z}(\beta,h)|_{h=0}.
\end{equation}
Further we will need the traces of powers of $H_0$, $
t_n\equiv  \Tr (H_0^n), n=0,1,2,\ldots, dim,$
and the set of traces
\begin{equation}\label{6}
{\cal T}(H_0)=\left\{ t_n|n=0,1,2,\ldots,dim\right\}.
\end{equation}
Two Hamilton operators with the same $N$ and $s$, $H_0^{(1)}=\bi{J}^{(1)}\cdot \bi{H}$ and
$H_0^{(2)}=\bi{J}^{(2)}\cdot \bi{H}$ are called \underline{isospectral} if they have the same
eigenvalues, counted with multiplicity, or, equivalently, if they generate the same characteristic
polynomial:
\begin{equation}\label{7}
\det(H_0^{(1)}-\lambda)=\det(H_0^{(2)}-\lambda)\quad \forall \lambda\in{\Bbb R}.
\end{equation}
According to the above remarks, we will also speak of ``isospectral systems".
Sometimes we will apply the term ``isospectral" to more general pairs of
operators derived from $H_0^{(i)}, i=1,2$ if there is no danger of confusion.
Clearly, (\ref{7}) defines an equivalence relation $\sim$ on ${\cal J}$.
The coefficients of a characteristic polynomial $P(\lambda)=\det(H_0-\lambda)$
can be viewed as polynomials of the coupling constants $J_{xy}$.
These polynomials assume constant values exactly on the $\sim$-equivalence classes $[\bi{J}]_\sim$,
which we will call \underline{isospectral classes}. Consequently, the isospectral classes are
\underline{algebraic varieties} in ${\cal J}$, since they are defined by a finite number $k$
of polynomial equations $p_\nu(\bi{J})=0, \nu=1,\ldots,k$.\footnote{
For an elementary introduction into the theory of algebraic varieties, see e.~g.~\cite{CLS}.
Note that algebraic varieties need not be differentiable manifolds since they
may contain ``boundaries" like the vertex of the light cone $x^2+y^2+z^2=c^2 t^2$
or ``hairs" like $x=y=0$ in the variety $(x^2+y^2)z=0$.}\\

In algebraic geometry there are various equivalent definitions of
the \underline{dimension} of an algebraic variety, which are, however, too technical to
be reproduced here (see, for example, \cite{CLS}, chapter 9).
For our purposes it will suffice to note that for the special case
where the Jacobian matrix
$\left( \frac{\partial p_\nu}{\partial J_j}\right)_{\nu=1\ldots k,j=1\ldots {N \choose 2}}$
has locally a constant rank $r$, the corresponding isospectral classes are locally
differentiable manifolds of dimension  $\ell={N \choose 2}-r$. This is an immediate
consequence of the fibration theorem (see, for example, \cite{AMR}, theorem 3.5.18).
Especially, if $k\ge {N \choose 2}$
and the rank of the Jacobian is maximal, $r={N \choose 2}$, the isospectrality
classes will only consist of a discrete point set.
Note, however, that the equivalence classes will never be trivial since they
at least consist of the orbits
of the group of discrete symmetries of ${\cal J}$
generated by permutations of spin sites.\\

We will say that the pair $(N,s)$ is of \underline{isospectral type} if the 
corresponding space of coupling constants ${\cal J}$
contains at least one isospectral equivalence class 
of dimension $\ell \ge 1$. The largest dimension $\ell$ of isospectral equivalence classes 
will be called the \underline{dimension} of the isospectral type.\\

Functions $f:{\cal J} \longrightarrow {\Bbb R}$ which are constant on isospectral classes will be called
\underline{isospectral invariants}. \\

Moreover, we will consider a special case of isospectrality in which the equivalence classes 
are easily calculated. Obviously, two isospectral Hamiltonians $H_0^{(1)}$ and $H_0^{(2)}$ can be related 
by a unitary transformation $U$:
\begin{equation}\label{8}
H_0^{(2)}=U^\ast H_0^{(1)} U.
\end{equation}
$U$ maps the eigenvectors of $H_0^{(2)}$ onto the corresponding eigenvectors of $H_0^{(1)}$.
In the case of 
a one--dimensional isospectral equivalence class parametrized by a coordinate $t$, $U$ can be 
chosen to depend smoothly on $t$. Following a closed loop, the corresponding unitary 
transformation need not reduce to the identity transformation, but may include some 
phases. We will come back to this phenomenon later.
If for sufficiently many curves of isospectrally equivalent points the corresponding unitary 
transformations $U(t)$ can, moreover, be chosen to be a one-parameter group, the system $(N,s)$ 
will be called of ``covariant isospectral" type. More precisely, we define:
\begin{defi}
$(N,s)$ is of \underline{covariant isospectral} type iff it is of isospectral type and for 
any isospectrally equivalent $\vec J^{(1)},\vec J^{(2)}$ which are sufficiently close, there
exists an anti-Hermitean  $dim\times$dim-matrix $\Omega$, and a real
${N\choose 2}\times{N\choose 2}$-matrix $M$ 
and some $t_0\in{\Bbb R}$, such that for all $t\in{\Bbb R}$
\begin{equation}\label{9}
\exp({\Omega^\ast t})\bi{J}^{(1)}\cdot\bi{H}\exp({\Omega t}) =
(\exp({-M t})\bi{J}^{(1)})\cdot\bi{H}
\end{equation}
and
\begin{equation}\label{10}
\exp({-M t_0})\bi{J}^{(1)}=\bi{J}^{(2)}.
\end{equation}
\end{defi}
It follows immediately, that the orbit   $\exp({-M t})\bi{J}^{(1)}$ in ${\cal J}$--space 
corresponds to a family of isospectral Hamilton matrices. Since in our case all Hamilton 
matrices are real and symmetric, $\Omega$ can also be chosen as real and hence anti--
symmetric, and thus $\exp({\Omega t})$ represents a rotation in the real Hilbert space ${\Bbb R}^{dim}$.\\

The condition (\ref{9}) may be replaced by its equivalent infinitesimal version:
\begin{equation}\label{11}
\left[ \bi{H},\Omega \right] = -M^\ast \bi{H}.
\end{equation}
This equation could be used to show that $M$, considered as a linear transformation in the 
space of ${N\choose 2}$--dimensional vectors with matrix entries endowed with the scalar 
product $\Tr (\bi{K}\cdot\bi{H})$, will be an antisymmetric matrix. We will give an 
independent proof of this fact later.\\

Further we note that the set of solutions $\langle \Omega,M \rangle$ of (\ref{11}) will be a 
Lie algebra with respect to the obvious vector and commutator operations. Hence covariant 
isospectral equivalence classes will be orbits of the corresponding matrix Lie groups.\\

Thus far we have only discussed isospectral systems in the
absence of an external magnetic field. If we include the field ${\cal H}$ and allow for a
corresponding $h$-dependence of the Hamiltonian, we have to consider a
slightly stronger concept of ``complete isospectrality".

\begin{defi}
Two Hamiltonians $H_0^{(1)}$ and $H_0^{(2)}$ (or, equivalently, two vectors $\bi{J}^{(1)}$
and $\bi{J}^{(2)}\in{\cal J}$) are called \underline{completely isospectral}
($\bi{J}^{(1)}\approx\bi{J}^{(2)}$)
iff for all $h\in{\Bbb R}\quad H^{(1)}(h)=H_0^{(1)}-h S^{(3)}$ and
$H^{(2)}(h)=H_0^{(2)}-h S^{(3)}$ are isospectral.
\end{defi}

\begin{prop}\label{P0}
The following conditions are equivalent: \\
\begin{tabular}{ll}
(i)  & $ \bi{J}^{(1)}\approx\bi{J}^{(2)}$,\\
(ii) & ${\bf P}_M H_0^{(1)}{\bf P}_M \sim {\bf P}_M H_0^{(2)}{\bf P}_M$\\
 & for all projectors ${\bf P}_M$ onto the eigenspaces of $S^{(3)}$ \\
 & corresponding to the eigenvalue $M$,\\
(iii)& ${\Bbb P}_S H_0^{(1)}{\Bbb P}_S \sim {\Bbb P}_S H_0^{(2)}{\Bbb P}_S$\\
 & for all projectors ${\Bbb P}_S$ onto the eigenspaces of $\bi{S}^2{}$ \\
 & corresponding to the eigenvalue $S(S+1)$.
\end{tabular}
\end{prop}

{\bf Proof}: (i)$\Rightarrow$ (ii). Since all $H(h), h\in{\Bbb R},$ commute, there
exists a  system of joint eigenprojectors ${\cal P}_{\nu,M_\nu}$ such that
\begin{equation}\label{11a}
H(h)=\sum_{\nu,M_\nu} (\epsilon_\nu - h M_\nu){\cal P}_{\nu,M_\nu}.
\end{equation}
Hence
\begin{equation}\label{11b}
{\bf P}_M H(h){\bf P}_M = \sum_{\nu,M_\nu,M_\nu=M}
(\epsilon_\nu - h M_\nu){\cal P}_{\nu,M_\nu}.
\end{equation}
If $H^{(1)}(h)\sim H^{(2)}(h)$ for all  $h\in{\Bbb R}$, then both systems have the same
set of eigenvalues $\epsilon_\nu - h M_\nu$  with the same multiplicities
$\Tr{\cal P}_{\nu,M_\nu}$. Hence, by (\ref{11b}), also
${\bf P}_M H^{(1)}(h){\bf P}_M \sim {\bf P}_M H^{(2)}(h){\bf P}_M$, especially
${\bf P}_M H_0^{(1)}{\bf P}_M \sim {\bf P}_M H_0^{(2)}{\bf P}_M$ for all $M$.\\

(ii)$\Rightarrow$(iii). The eigenspaces corresponding to ${\cal P}_{\nu,M_\nu}$ can be
further split into eigenspaces of $\bi{S}^2$. This can be written as
\begin{equation}\label{11c}
{\cal P}_{\nu,M_\nu} =\sum_{S=|M_\nu|}^{S_{max}}{\cal P}_{\nu,M_\nu,S}.
\end{equation}
Applying the ladder operators $ S^\pm$ gives
\begin{equation}\label{11d}
\Tr{\cal P}_{\nu,M_\nu,S} =\Tr{\cal P}_{\nu,S,S}=\Tr{\cal P}_{\nu,S}
\mbox{ for all }M_\nu=-S,\ldots,S.
\end{equation}
These numbers are the multiplicities of the eigenvalues of the operators
${\Bbb P}_S{\bf P}_M H_0^{(i)}{\Bbb P}_S {\bf P}_M$,
which are hence isospectral for $i=1,2$. By summation over $M$ we conclude that
also ${\Bbb P}_S H_0^{(i)}{\Bbb P}_S $,
are isospectral for $i=1,2$.\\

(iii)$\Rightarrow$(ii). This can be shown analogously by considering
\begin{equation}\label{}
{\Bbb P}_S H_0{\Bbb P}_S =\sum_{\nu,S_\nu,S_\nu=S} \epsilon_\nu {\cal P}_{\nu,S_\nu}',
\end{equation}
\begin{equation}\label{}
{\cal P}_{\nu,S_\nu}'=\sum_{M=-S_\nu}^{S_\nu} {\cal P}_{\nu,S_\nu,M}'
\end{equation}
and
\begin{equation}\label{}
\Tr{\cal P}_{\nu,S_\nu}'=(2 S_\nu +1) \Tr {\cal P}_{\nu,S_\nu,M}' .
\end{equation}

(ii)$\Rightarrow$(i). This follows from
\begin{equation}\label{}
H(h) = \sum_M ({\bf P}_M H_0 {\bf P}_M -h M).
\end{equation}
\hspace*{\fill}\rule{3mm}{3mm}

Obviously, $\bi{J}^{(1)}\approx\bi{J}^{(2)}$ implies $\bi{J}^{(1)}\sim\bi{J}^{(2)}$,
but not conversely:\\
Take $N=3, s=\frac{1}{2}$, then $\bi{J}^{(1)}=\left(\begin{array}{c} 1\\1\\1\end{array}\right)\sim
\bi{J}^{(2)}=\left(\begin{array}{c} -1\\-1\\-1\end{array}\right)$,
but $\bi{J}^{(1)}\not\approx\bi{J}^{(2)}$.\\
Unfortunately we do not know of less trivial counter-examples.
The problem is the following:\\
Being rotationally symmetric,
$H_0$ commutes with $\bi{S}^2$ and $S^{(3)}$, hence each eigenspace of $H_0$ with eigenvalue
$\epsilon_\nu$ is spanned by simultaneous eigenvectors of $\bi{S}^2$ and $S^{(3)}$, say,
$|\nu,\lambda,\mu\rangle,
\nu=0,\ldots,D, \lambda=1,\ldots,d_\nu, \mu=-S_{\nu,\lambda},\ldots,S_{\nu,\lambda}$, such that
\begin{equation}\label{13}
\bi{S}^2 |\nu,\lambda,\mu\rangle = S_{\nu,\lambda}(S_{\nu,\lambda}+1)|\nu,\lambda,\mu\rangle,
\quad
S^{(3)}|\nu,\lambda,\mu\rangle = \mu|\nu,\lambda,\mu\rangle.
\end{equation}
The degeneracy of the eigenvalues $\epsilon_\nu$ will be
\begin{equation}\label{13a}
n_\nu =\sum_{\lambda=1}^{d_\nu}(2 S_{\nu,\lambda}+1).
\end{equation}
We do not assume that the $S_{\nu,\lambda}$ have different values.
The corresponding eigenvalues of $H(h)=H_0-hS^{(3)}$ are
\begin{equation} \label{14}
E_{\nu,\lambda,\mu}= \epsilon_\nu - h \mu.
\end{equation}
For ``generic" $\bi{J}\in{\cal J}$ we expect that $H_0$ will have no further degeneracy besides
that dictated by rotational symmetry (``minimal" degeneracy), i.~e.~we expect that $d_\nu=1$.
So $\lambda$ can be skipped and $\mu=-S_\nu,\ldots,S_\nu$.
In this case, ``isospectrality" and ``complete isospectrality" would be equivalent. So, heuristically,
we may consider these two notions as having equal meaning, although we have to distinguish between them
for the sake of mathematical rigour.\\

For the most important case of isospectral systems which are obtained by continuously
varying the coupling constants we can show the following:
\begin{prop}\label{P01}
Any two systems joined by a continuous curve of isospectral systems are completely
isospectral.
\end{prop}
{\bf Proof}: Consider a curve $t\mapsto\bi{J}(t)$ and consider
\begin{equation}\label{11e}
H_0(t)\equiv \bi{H}\cdot \bi{J}(t) = \sum_\nu \epsilon_\nu {\cal P}_\nu(t)
\end{equation}
and
\begin{equation}\label{11f}
{\Bbb P}_S H_0(t){\Bbb P}_S = \sum_\nu \epsilon_\nu {\cal P}_\nu(t) {\Bbb P}_S.
\end{equation}
Since $t\mapsto{\cal P}_\nu(t)$ is continuous and $\Tr ({\cal P}_\nu(t) {\Bbb P}_S)$
assumes only non-negative integer values,
the latter must be constant w.~r.~t.~the parameter $t$. However,
$\Tr ({\cal P}_\nu(t) {\Bbb P}_S)$ equals the multiplicity of the eigenvalue $\epsilon_\nu$
in (\ref{11f}). Thus all ${\Bbb P}_S H_0(t){\Bbb P}_S, t\in{\Bbb R},$ are isospectral,
and, by proposition \ref{P0}, all $H_0(t)$ are completely isospectral.
\hspace*{\fill}\rule{3mm}{3mm}

Note that, according to this proposition,
in the above counter-example the two systems cannot lie in the same
connected component of an isospectral class.\\

We add some definitions concerning symmetrical polynomials which will be of later use:\\
The equation
\begin{equation}\label{15}
\prod_{n=1}^{d}{(x+x_n)}=\sum_{\nu =0}^{d}{s_\nu  x^{d-\nu }}
\end{equation}
defines the elementary symmetrical polynomials
\begin{equation}\label{16}
s_1=\sum_{n=1}^{d}{x_n},\quad s_2=\sum_{n<m}^{}{x_n x_m},
\quad\ldots\quad s_d=\prod_{n=1}^{d}{x_n}.
\end{equation}
These also appear, up to a sign, as the coefficients of the characteristic polynomial of $H_0$
\begin{equation}\label{17}
p(\lambda) = \prod_{i=1}^{d}{(\lambda-E_i)},
\end{equation}
written as polynomials of the $E_i$, where $d=dim$.
Every other symmetric polynomial can uniquely be written
as a polynomial of the $s_\nu $ (see, for example \cite{CLS}, 7.1, Theorem 3). This holds especially for
\begin{equation}\label{18}
t_n=\Tr (H_0^n)=\sum_{i=1}^{d}{E_i^n}.
\end{equation}
Conversely, each $s_\nu$ can uniquely be written as a polynomial of the $t_n, n=1\ldots d$,
(see, for example \cite{CLS}, 7.1, Theorem 8)
e.~g.~
\begin{equation}\label{19}
s_5=\frac{1}{5!}\left(t_1^5- 10 t_1^3 t_2 + 15 t_1 t_2^2 + 20 t_1^2 t_3 - 20 t_2 t_3 -
30 t_1 t_4 + 24 t_5 \right).
\end{equation}
This representation is independent of the dimension $d$.

\section{Specific heat and magnetic susceptibility}

As mentioned in the introduction, (completely) isospectral spin systems will give rise to the same
thermodynamic functions like specific heat and magnetic susceptibility. In this section we will
state this more precisely and also
prove the converse, up to the subtle distinction between complete and plain isospectrality.

\begin{lemma}\label{L1}
Two spin systems are isospectral iff ${\cal T}(H_0^{(1)})={\cal T}(H_0^{(2)})$.
\end{lemma}
{\bf Proof}:
Recall that ${\cal T}(H_0)$ was defined as the set of traces $t_n=\Tr (H_0^n), n=0,\ldots,dim$.
Hence the ``only if" part is obvious. From the remarks at the end of section 2 it follows that
if two systems possess the same $t_n, n\in{\Bbb N}$,
they share also the same values of the standard symmetric polynomials
$s_\nu(\bi{E})$ and hence have the same characteristic polynomial $p(\lambda)$.
\hspace*{\fill}\rule{3mm}{3mm}

\begin{prop}\label{P1}
Two spin systems are isospectral iff they possess the same specific heat function.
\end{prop}
{\bf Proof}: $c(\beta )$ can be expanded into a Taylor series at $\beta =0$:
\begin{eqnarray}\label{20}
c(\beta ) & = & \left(-\frac{t_1^2}{t_0^2}+\frac{t_2}{t_0} \right)\beta^2\\  \nonumber
 & & +\left(\frac{t_1 t_2}{t_0^2}-\frac{t_1}{t_0}\left(\frac{t_1^2}{t_0^2}-
\frac{t_2}{2 t_0}\right)-\frac{t_3}{2 t_0} \right)\beta^3\\ \nonumber
& & +\ldots .
\end{eqnarray}
This is the starting point of the so-called  ``moment expansion method", see e.~g.~\cite{Yos} 7.3.
Obviously, each coefficient of $\beta^n$ uniquely determines $t_n$, if the other traces
$t_m, m<n,$ are already known. Note that $t_0=dim, t_1=0$. Together with lemma \ref{L1}
this completes the proof.\hspace*{\fill}\rule{3mm}{3mm}

Now we consider again $H(h)=H_0-h S^{(3)}$ with eigenvectors $|\nu,\lambda,\mu\rangle$ according to the
previous section. By its very definition, two completely isospectral systems
share the same partition function ${\cal Z}(\beta,h)$ and any other thermodynamic function which
can be derived from it. Especially, the following holds:

\begin{prop}\label{P11}
Two completely isospectral systems possess the same magnetic susceptibility function.
\end{prop}

To tackle the converse problem, we consider
${\cal Z}(\beta,0)=\sum_\nu n_\nu \exp({-\beta\epsilon_\nu})$ and
define the coefficients $\sigma_\nu$ implicitely by
\begin{equation}\label{21}
\Tr \left( S^{(3)2} \exp({-\beta H_0}) \right) = \sum_\nu \sigma_\nu n_\nu
\exp({-\beta\epsilon_\nu})
\end{equation}
In the case of minimal degeneracy, i.~e.~$d_\nu=1$, we have $n_\nu= 2 S_\nu +1$ and
\begin{equation}\label{22}
\sigma_\nu=\frac{1}{2 S_\nu +1}\sum_{\mu=-S_\nu}^{S_\nu}\mu^2 =\frac{1}{3}S_\nu (S_\nu +1).
\end{equation}
In the general case,
\begin{eqnarray} \label{23}
\sigma_\nu & = &  \frac{1}{n_\nu }\sum_{\lambda=1}^{d_\nu}
\sum_{\mu=-S_{\nu,\lambda}}^{S_{\nu,\lambda}} \mu^2 \\ \nonumber
&=&\frac{1}{n_\nu }\sum_{\lambda=1}^{d_\nu}\frac{1}{3}S_{\nu,\lambda}
 (S_{\nu,\lambda} +1)(2 S_{\nu,\lambda}+1).
\end{eqnarray}
\begin{prop}\label{P2}
Two spin systems with the same susceptibility function are  isospectral.
\end{prop}
{\bf Proof}: Since
\begin{equation}\label{24}
\chi(\beta)=\frac{\beta}{{\cal Z}(\beta,0)}\Tr  \left( S^{(3)2}\exp({-\beta H_0}) \right),
\end{equation}
the two systems will have the same function
\begin{eqnarray}\label{25}
f(\beta) & \equiv  & \frac{\Tr \left( \exp({-\beta H_0}) S^{(3)2} \right) }
{\Tr \left(\exp({-\beta H_0})\right)} \\  \nonumber
 & = & \frac{\sum_{\nu=0}^{D}{\sigma_\nu n_\nu \exp({-\beta \varepsilon_\nu})}}
 {\sum_{\nu=0}^{D}{n_\nu \exp({-\beta \varepsilon_\nu})}}.
\end{eqnarray}
Since $\varepsilon_0 < \varepsilon _1 < \varepsilon _2 < \ldots $, the terms
$\exp({-\beta \varepsilon_\nu})$ are of different orders of magnitude for
$\beta \rightarrow \infty $. The first term increasingly dominates, hence
\begin{equation}\label{26}
\lim_{\beta \to \infty }f(\beta) = \frac{\sigma _0 n_0 \exp({-\beta\varepsilon_0 })}
{n_0 \exp({-\beta\varepsilon_0})} = \sigma _0.
\end{equation}
If we subtract this limit from $f(\beta)$, the dominant term asymptotically becomes
\begin{eqnarray}\label{27}
f(\beta)-\sigma _0 & \simeq _{\beta\to\infty} & \frac{\sigma_1 n_1
\exp({-\beta\varepsilon_1 })}
{ n_0 \exp({-\beta\varepsilon_0 })}\\  \nonumber
 & = & \frac{\sigma_1 n_1}{n_0}\exp({-\beta(\varepsilon_1 -\varepsilon_0 ))}.
\end{eqnarray}
In the next step we have
\begin{equation}\label{28}
f(\beta)-\sigma_0 -\frac{\sigma_1 n_1}{n_0}\exp({-\beta(\varepsilon_1 -\varepsilon_0 )})
\simeq _{\beta\to\infty}
\frac{\sigma _2 n_2}{n_0} 2\exp({-\beta(\varepsilon_2 -\varepsilon_0 )}),
\end{equation}
and so on.
In this way, from the behaviour of $f(\beta)$ for $\beta\to\infty$, we may extract the values\\
$\sigma _0,\frac{\sigma_1 n_1}{n_0},\ldots  \frac{\sigma_\nu n_\nu}{n_0},\ldots
\frac{\sigma_D n_D}{n_0}$
and
$\varepsilon_1 - \varepsilon_0, \varepsilon_2 -\varepsilon_0,
\ldots, \varepsilon_\nu - \varepsilon_0, \ldots, \varepsilon_D - \varepsilon_0$.\\
Let $t_n = \Tr (H_0^n)$ as above and $\mu_n\equiv  \Tr (H_0^n S^{(3)2}), n\in{\Bbb N} $.
$\mu_0 = \Tr (S^{(3)2})$ can be calculated independently of $H_0$. Since
\begin{equation}\label{29}
\mu_0=\sum_{\nu=0}^{D}{\sigma_\nu n_\nu} = n_0 \sum_{\nu=0}^{D}{\frac{\sigma_\nu n_\nu}{n_0}} ,
\end{equation}
$n_0$ and hence $\sigma _\nu n_\nu, \nu=1\ldots D,$ are also uniquely determined.\\
Next we consider the Taylor expansion of $f(\beta)$ at $\beta=0$:
\begin{eqnarray}  \label{30}
f(\beta) & = & \frac{\mu_0}{t_0}-\frac{\mu_1}{t_0}\beta +
\left(-\frac{t_2 \mu_0}{2 t_0^2}+\frac{\mu_2}{2 t_0}\right)\beta^2\\   \nonumber
& & +\left(\frac{t_3 \mu_0}{6 t_0^2}+\frac{t_2 \mu_1}{2 t_0^2}-\frac{\mu_3}{6 t_1}\right)\beta^3\\  \nonumber
& & +\left(\mu_0\left(\frac{t_2^2}{4 t_0^3}-\frac{t_4}{24 t_0^2}\right)
-\frac{t_3 \mu_1}{6 t_0^2}-\frac{t_2 \mu_2}{4 t_0^2}+\frac{\mu_4}{24 t_0}\right)\beta^4\\  \nonumber
& & +\ldots
\end{eqnarray}
Recall that $t_0, \mu_0$ are known. The linear term then gives
\begin{equation}\label{31}
\mu_1=\Tr (H_0 S^{(3)2})=\sum_{\nu_0=0}^{D}{\sigma_\nu n_\nu \varepsilon_\nu}.
\end{equation}
On the other side we know the l.~h.~s.~of
\begin{equation}\label{32}
\sum_{\nu=0}^{D}{\sigma_\nu n_\nu (\varepsilon_\nu-\varepsilon_0 )} = \mu_1 - \varepsilon_0 \mu_0,
\end{equation}
hence $\varepsilon_0$ and all $\varepsilon_\nu, \nu=1\ldots D$ are also known. Similarly,
\begin{equation}\label{33}
\sum_{\nu=0}^{D}{\sigma_\nu n_\nu (\varepsilon_\nu-\varepsilon_0 )^2} =
\mu_2 - 2\varepsilon_0 \mu_1 + \varepsilon_0^2\mu_0 ,
\end{equation}
hence $\mu_2$ is known and from the $\beta^2$--term in (\ref{30})  also $t_2$, and so on.\\
Eventually, we obtain all $t_n, \mu_n, n=2\ldots dim$ solely from  $\chi(\beta)$.
According to lemma \ref{L1}
this gives us all eigenvalues of $H_0$ with multiplicity, i.~e.~$n_\nu, \nu=0,\ldots D$
and the two spin systems are isospectral. \hspace*{\fill}\rule{3mm}{3mm}

The proof does not give complete isospectrality: If some eigenvalue $\varepsilon_\nu$
belongs to different $S_{\nu\lambda}, \lambda > 2$ then from
$n_\nu=\sum_{\lambda}^{}{(2S_{\nu\lambda}+1)}$ and
$\sigma_\nu = \frac{\sum_{\lambda}^{}{\frac{1}{3}
S_{\nu\lambda}(S_{\nu\lambda}+1)(2S_{\nu\lambda}+1)}}
{\sum_{\lambda}^{}{(2 S_{\nu\lambda}+1)}}$ the $S_{\nu,\lambda}$ cannot be uniquely determined.

\section{Some isospectral invariants}

Criteria for non--isospectrality could, in principle, be checked by brute force methods: 
Calculate the characteristic polynomial of the matrix $H_0=\bi{J}\cdot\bi{H}$, say
$p(\lambda)=\sum_{\nu=0}^{dim} c_\nu \lambda^\nu$. Select ${N\choose 2}$ different 
coefficients $c_i, c_j,\dots$ ($\nu=dim$ being excluded since $c_{dim}=1$) and calculate the 
Jacobian determinant
\begin{equation}\label{34}
Jac(\bi{J})\equiv  \frac{\partial(c_i,c_j,\ldots)}{\partial(J_1,\ldots,J_{N\choose 2})}
\end{equation}
preferably by using a computer algebra software.
If the Jacobian nowhere vanishes, according to remarks in section 2,
$(N,s)$ cannot be of isospectral type.\\

In practice this method will, even for small $N$ and $s$, rapidly become extremely memory- 
and time-consuming. To simplify the problem one could -- in the case of complete isospectrality --
restrict oneself to subspaces invariant 
under $H_0$, for example subspaces ${\cal H}(M)$ of constant magnetic quantum number $M$.\\

The space with maximal $M$, ${\cal H}(M = N s)$ is one-dimensional and is spanned by the 
product state
\begin{equation}\label{35}
\varphi_0=|s,s,\ldots s\rangle
\end{equation}
which is an eigenstate of $\bi{J}\cdot\bi{H}$ for all $\bi{J}\in {\cal J}$ with eigenvalue
\begin{equation}\label{36}
E_0=s^2 J \equiv  s^2 \sum_{x<y}J_{xy}.
\end{equation}
This proves the first part of the following
\begin{lemma}\label{L2}
If $\bi{J}^{(1)}\cdot\bi{H}$ and $\bi{J}^{(2)}\cdot\bi{H}$ are completely or covariantly isospectral,
then
\begin{equation}\label{37}
\sum_{x<y}J_{xy}^{(1)} = \sum_{x<y}J_{xy}^{(2)},
\end{equation}
i.~e.~$\bi{J}^{(1)}$ and $\bi{J}^{(2)}$ lie in the same hyperplane perpendicular to 
$\boldsymbol{1}\equiv (1,1,\ldots 1)$.
\end{lemma}
{\bf Proof}: If all $J_{xy}\ge 0$, then $E_0=s^2 J$ will be the maximal eigenvalue of
$\bi{J}^{(2)}\cdot\bi{H}$.
In fact, $\langle \varphi_0| H_{xy} \varphi_0 \rangle$ is the maximal expectation value
for each $H_{xy}$. Hence $J$ is an isospectral invariant at least in the domain
${\cal J}^+\equiv \{\bi{J}|\mbox{ all } J_{xy}\ge 0\}$. Now assume covariant isospectrality and let
$t\mapsto \bi{J}(t)=\exp({-M t}) \bi{J}(0)$ be an isospectral curve which will be restricted to
${\cal J}^+$. According to what has been said before,
$0=\dot{J}=\frac{d}{dt}\langle \bi{J} |\boldsymbol{1}\rangle = \langle \dot{\bi{J}} |\boldsymbol{1}\rangle
= \langle -M \bi{J} |\boldsymbol{1}\rangle = \langle \bi{J} |-M^\ast \boldsymbol{1}\rangle
=\langle \bi{J} |M \boldsymbol{1}\rangle$ for all $\bi{J}\in{\cal J}^+$. Since ${\cal J}^+$ linearly
generates ${\cal J}$ it follows that $M\boldsymbol{1}=\boldsymbol{0}$, i.~e.~all row sums of $M$ vanish
and $J$ is an isospectral invariant on the whole space ${\cal J}$.
\hspace*{\fill}\rule{3mm}{3mm}

Before proceding with $M=Ns-1$ we will show that also
\begin{lemma}\label{L4}
$||\bi{J}||^2=\sum_{x<y}J_{xy}^{2}$ is an isospectral invariant.
\end{lemma}
{\bf Proof}: Obviously, $\Tr (H_0^2)$, the sum of all eigenvalues squared, is the same for isospectral 
Hamiltonians. After expanding the square $(\bi{J}\cdot\bi{H})^2$ one realizes that only those 
products $H_{xy}H_{uv}$ have a non-zero trace where $x=u$ and $y=v$. Hence
$\Tr (H_0^2)= \sum_{x<y}J_{xy}^2 \Tr (H_{xy}^2)=(\sum_{x<y}J_{xy}^2 )\cdot 
\frac{1}{3}s^2(s+1)^2(2s+1)^N$, see Appendix A. The actual value for $\Tr (H_{xy}^2)$ 
is irrelevant for the proof; what matters only is that it is independent of $x,y$.
This concludes the proof.
\hspace*{\fill}\rule{3mm}{3mm} \\

From Lemma \ref{L4} it follows immediately that in the covariant isospectral case the matrix
$\exp({-t M})$ leaves the Euclidean norm $||\ldots||$ invariant and thus must be an orthogonal 
transformation and its generator $-M$ will be antisymmetric.

\section{The triangle ($N=3$)}

\subsection{$s=\frac{1}{2}$}

Next we consider eigenvalues with eigenvectors in the subspace ${\cal H}(M=Ns-1)$, but 
restricted to the case $N=3$. With the abbreviations
\begin{equation}\label{38}
J=J_{12}+J_{23}+J_{13},
\end{equation}
\begin{equation}\label{39}
\Gamma=J_{12}J_{23}+J_{12}J_{13}+J_{23}J_{13},
\end{equation}
the eigenvalues are calculated to be (see \cite{LL}, {\S}  62, Exercise 2 )
\begin{equation}\label{40}
E_0= s^2 J,\quad   E_{1,2}=s(s-1)J\pm \sqrt{J^2-3 \Gamma}.
\end{equation}
Hence the first three eigenvalues are constant on curves with constant $J$ and $\Gamma$.
These are circles with radius
\begin{equation}\label{41}
r=\sqrt{\frac{2}{3}J^2-2\Gamma},
\end{equation}
the center of which is located on the line $J_{12}=J_{23}=J_{13}$, including the degenerate 
case $r=0$.\\

For $s=\frac{1}{2}$ the list of eigenvalues is already exhausted: Due to rotational symmetry 
the value $E_0$ is $4$--fold degenerate $(S=\frac{3}{2})$ and the $E_{1,2}$ are two--fold 
degenerate $(S=\frac{1}{2})$. We conclude:
\begin{prop}\label{P3}
The system $N=3, s=\frac{1}{2}$ is of complete isospectral type with dimension $2$.
\end{prop}

We now consider the question whether the triangle with $s=\frac{1}{2}$ is of covariant 
isospectral type, i.~e.~we seek solutions of
\begin{equation}\label{43}
[\bi{H},\Omega]= M \bi{H},
\end{equation}
$\Omega$ and  $M$ being anti--symmetric. Let $T=T^3$ be the unitary left shift operator which
represents a cyclic permutation of the spin sites. Then a 
solution of (\ref{43}) is given by
\begin{equation}\label{44}
\Omega = \frac{1}{2\sqrt{3}}(T-T^\ast),
\quad M=\frac{1}{\sqrt{3}} \left( \begin{array}{ccc}0&1&-1\\-1&0&1\\1&-1&0 \end{array}\right).
\end{equation}
$\Omega$ can also be written as
\begin{equation}\label{45}
\Omega=\frac{i}{4\sqrt{3}}\boldsymbol{\sigma}_3\cdot
(\boldsymbol{\sigma}_1\times\boldsymbol{\sigma}_2),
\end{equation}
where the $\boldsymbol{\sigma}_i, (i=1,2,3)$ denote the Pauli-matrices.
Obviously, $\Omega$ is rotationally symmetric
which entails complete isospectrality.
The factor $\frac{1}{\sqrt{3}}$ is chosen such that the parameter $t$ in $\exp({tM})$ will be just 
the angle of rotation. $T^3=T$ entails $\Omega^3=-\frac{1}{4}\Omega$, hence the exponential 
series of $\exp({t\Omega})$ will be actually a polynomial in $\Omega$:
\begin{equation}\label{46}
\exp({t\Omega)}=1+2\Omega\sin\frac{t}{2}+(2\Omega)^2(1-\cos\frac{t}{2}).
\end{equation}
For special values of $t$ we obtain:
\begin{equation}\label{47}
\exp({\frac{4\pi}{3}\Omega})=T,
\end{equation}
\begin{equation}\label{48}
\exp({4\pi\Omega})=T^3=\Eins,
\end{equation}
\begin{equation}\label{49}
\exp({2\pi\Omega})=\frac{2}{3}(T^2+T-\frac{1}{2}).
\end{equation}
The last expression can be rewritten using
\begin{equation}\label{50}
\tilde{H}\equiv \left( \begin{array}{c}1\\1\\1\end{array} \right)\cdot\bi{H}=
\frac{3}{4}\left( {\Bbb P}_{\frac{3}{2}} -{\Bbb P}_{\frac{1}{2}} \right),
\end{equation}
where ${\Bbb P}_{\frac{3}{2}}$ (resp.~${\Bbb P}_{\frac{1}{2}}$) denotes the projector onto the 
subspace $S=\frac{3}{2}$ (resp.~$S=\frac{1}{2}$). The result is
\begin{equation}\label{51}
\exp({2\pi\Omega})={\Bbb P}_{\frac{3}{2}} -{\Bbb P}_{\frac{1}{2}}.
\end{equation}
This means that an eigenstate of $H_0$ with $S=\frac{1}{2}$ acquires a phase of $\pi$
after a full rotation in ${\cal J}$--space, analogous to the occurence of Berry phases
for adiabatic loops in parameter space. Summarizing, we state the following proposition
which, in essence, is due originally to V.~G.~Grachev \cite{G1}:
\begin{prop}\label{P5}
The system $N=3, s=1/2$ is of both completely and covariantly isospectral type.
\end{prop}

\subsection{A physical example}

An interesting and timely application of this theory is provided by the
example of the molecular
magnet (CN$_3$H$_6$)$_4$Na$_2$[H$_4$V$_6$P$_4$O$_{30}$(CH$_2$)$_3$CCH$_2$OH$_2$]$\bullet$ 14H$_2$O
which features two uncoupled systems of three $V^{4+}\quad (s =\frac{1}{2})$
ions that interact via
antiferromagnetic Heisenberg exchange. It has been proposed \cite{MMBSB}
that the Coulomb interaction
between an Na ion and two of the three $V^{4+}$ ions gives rise to what is
essentially an isosceles
triangle, with the distances between the three vanadium ions being $3.20,
3.21,$ and $3.36 ${\AA}. It is
then quite reasonable to assume, that the three exchange constants
satisfy $J_{12} = J_{13} \neq J_{23}$ . In fact,
calculation of the weak field susceptibilty has yielded results that are
in excellent agreement with
accurate susceptibility measurements from room temperature down to $2 K$
upon assigning the
values $J_{12} = J_{13}= 64.7K$ and $J_{23} = 7.5 K$
\footnote{It is usual to measure the coupling constants in units of Kelvin. The
corresponding energies are obtained by multiplying with the Boltzmann constant $k_B$.}
\cite{Jun}, \cite{Lub}.
Moreover, the
calculated energy level spacings
that follow from these assignments have recently been confirmed to good
accuracy in a direct
manner by inelastic neutron scattering \cite{St}. Nevertheless, as the work
of this section has
shown, the identical energy levels and the identical temperature
dependent susceptibility emerge
for the continuous choices of the three different exchange constants
that lie on curves with
$J = 136.9K$ and $\Gamma = 5156.6 K^2$ .

\subsection{$s >\frac{1}{2}$}

For $s>\frac{1}{2}$ we consider the next subspace ${\cal H}(M=3s-2)$. The characteristic 
polynomial $p(\lambda)=\sum_{\nu=1}^6 c_\nu \lambda^\nu$ has been calculated using 
MATHEMATICA$^{\circledR}$ 4.0, but is too complicated to be presented here. One particular Jacobian reads
\begin{equation}\label{42}
\frac{\partial (c_3,c_4,c_5)}{\partial(J_{12},J_{23},J_{13})}=
(3s-1)(6s-1)(1-8s+6s^2)(1-6s+15s^2).
\end{equation}
This is a polynomial in $s$ which has no integer or half--integer roots. Therefore we have 
proved the following
\begin{prop}\label{P4}
The system $N=3, s>\frac{1}{2}$ is not of complete isospectral type.
\end{prop}

\section{Isospectrality for $N>3$}

The question arises whether our result that complete isospectrality only occurs for $s=\frac{1}{2}$
also holds for $N>3$. Our method of calculating the Jacobian (\ref{34}) for arbitrary $s$
will no longer work for larger $N$. However, we can prove a weaker statement, namely
\begin{prop} \label{P6}
Systems with $N>3$ cannot be of covariantly isospectral type.
\end{prop}
{\bf Proof:} For this we need a trace formula which  will 
be explained in the Appendix A:
\begin{eqnarray}\label{52}
\Tr (H_0^3) & = & \sum_{x<y} (J_{xy})^3\left( -\frac{1}{6} s^2(s+1)^2(2s+1)^N  \right)\\ \nonumber
        & + & \sum_{x<y<z}J_{xy}J_{yz}J_{xz} \left( \frac{2}{3} s^3(s+1)^3(2s+1)^N \right).
\end{eqnarray}
From the isospectral invariance of $\Tr (H^3)$ we conclude that also
\begin{equation}\label{53}
f_3\equiv \sum_{x<y} (J_{xy})^3 - 4s(s+1)\sum_{x<y<z}J_{xy}J_{yz}J_{xz}
\end{equation}
will be invariant. Now we consider four different spin sites (using $N>3$) denoted by $1,2,3,4$ and
consider vectors $\bi{J}\in{\cal J}$ which have vanishing components except possibly for
$J_{12},J_{13},J_{23},J_{14},J_{24},J_{34}$. One-parameter isospectral curves passing through
$\bi{J}$ satisfy
\begin{equation}\label{55}
\frac{d}{dt}f_3(\bi{J}(t)) = 0.
\end{equation}
Using (\ref{53}) and $\frac{d}{dt}\bi{J}(t)=-M \bi{J}(t)$, (\ref{55}) can be written as an equation
which is linear w.~r.~t.~the $15$ relevant matrix entries of $M$ and trilinear w.~r.~t.~the 6
non-vanishing components of $\bi{J}$. Using MATHEMATICA$^{\circledR}$ 4.0 it is easy to show that (\ref{55}) has only the trivial solution
$M=0$. For example, one may randomly choose 15 vectors $\bi{J}$ (it suffices to consider components
$-1,0,1$) and cast the corresponding $15 $ equations of the form (\ref{55}) into matrix form. Non-trivial
solutions exist only if the determinant of this matrix,
which is a polynomial in $s$ of degree 30 with integer coefficients, vanishes.
However, the zeros of the polynomial can be numerically computed and
shown not to attain half-integer or integer values.
\hspace*{\fill}\rule{3mm}{3mm}  \\

So it seems that the concept of covariant isospectrality is of little use
having only one single application for $N=3, s=1/2$. However, covariance may be restored for $N\ge 4$
if the class of admissible Hamiltonians is suitably extended, e.~g.~to include also Hamiltonians
which are bi-quadratic in the spin observables. However, this is beyond the scope of the present article.\\

Of course, our proposition \ref{P6} does not exclude plain isospectrality for $N>3$. Indeed, we 
will show that the system $N=4, s=\frac{1}{2}$ is completely isospectral, albeit not covariantly 
isospectral, in the next section.

\section{The $s=1/2$ tetrahedron case}

In the case $N=4, s=1/2$ it is still possible to calculate the coefficients of the 
characteristic polynomials of $H_0$ restricted to the subspaces with $M=0, S=2,1,0$.
Obviously, this is enough in order to study complete isospectrality
since all eigenvalues of $H_0$ appear within these subspaces.
It turns out that all coefficients can be written as functions of four fundamental
invariants $I_1,I_2,I_3,I_4$. These can most conveniently be
written in terms of new coordinates in ${\cal J}$, which
are defined as half the sums and differences of the  coupling
constants of adjacent edges:
\begin{equation} \label{56}
S_{12}\equiv  \frac{1}{2}(J_{12}+J_{34}),\quad D_{12}\equiv  \frac{1}{2}(J_{12}-J_{34}),
\end{equation}
etc.\\

\begin{prop}\label{P8}
Two spin systems with coupling constants
$\bi{J}^{(1)},\bi{J}^{(2)}$ with $N=4, s=1/2$ are completely isospectral
iff the following four  functions assume the same values for
$\bi{J}^{(1)}$ and $\bi{J}^{(2)}$:
\begin{equation} \label{57}
I_1=D_{12}^2+D_{13}^2+D_{14}^2,
\end{equation}
\begin{equation} \label{58}
I_2=S_{12}^2+S_{13}^2+S_{14}^2,
\end{equation}
\begin{equation} \label{59}
I_3=S_{12}+S_{13}+S_{14},
\end{equation}
\begin{equation} \label{60}
I_4=2 D_{12}D_{13}D_{14}+D_{12}^2 S_{12}+D_{13}^2 S_{13}+D_{14}^2 S_{14}-S_{12}S_{13}S_{14}.
\end{equation}
\end{prop}

Now let $I'$ be the functional matrix obtained by partial differentiation of $I_1,I_2,I_3,I_4$
with respect to its 6 arguments $S_{12},\cdots,D_{14}$.
The rank of $I'$ assumes its maximal value of 4 iff no determinant of the 15 possible
$4\times 4$ submatrices of $I'$ vanishes. We denote the subset of those points with
maximal rank by ${\cal R}\subset {\cal J}$.
If an isospectral class lies entirely within ${\cal R}$ it will be
a 2-dimensional submanifold of the 6-dimensional space ${\cal J}$. This follows
by a well-known theorem of differential geometry (see e.~g.~\cite{AMR} Theorem 3.5.4).
We will call this case \underline{generic}, the other cases \underline{exceptional}.\\

Although ${\cal J}$ is six-dimensional,
one can visualize the 2-dimensional submanifolds in the generic case.
$I_1=const.$ defines a sphere in the 3-dimensional $\bi{D}$-space with
coordinates $D_{12},D_{13},D_{14}$. $I_2=const.$ and
$I_3=const.$ define the intersection of a sphere and a plane,
i.~e.~a circle in $\bi{S}$-space.
For given $\bi{D}$ the last equation $I_4(\bi{D},\bi{S})=const.$
picks out a finite number (actually $\le 6$) of points in that circle.
If the corresponding angles $\psi_\nu$ are
drawn as different radii $r_\nu = 1+\frac{\psi_\nu}{4 \pi}$ in
$\bi{D}$-space (identifying points with $r_\nu=1/2 \mbox{ and } 3/2$)
we obtain a surface folded in a complicated way.\\

A large number of 2-dimensional generic isospectral classes have been identified numerically.
The exceptional classes are one- or zero-dimensional and will be
discussed further in  Appendix B.

\section{Heuristic arguments for the (non-)occurrence of isospectrality}

In the previous sections we have studied isospectrality for the cases
$s=\frac{1}{2}, N=3,4$ and excluded certain other cases, e.~g.~complete isospectrality
for $s>\frac{1}{2}$ and covariant isospectrality for $N>3$. However, we have been unable to present
a complete list of criteria for the (non-)occurrence of isospectrality. What is also
missing is some simple and intuitive argument
why isospectrality is so rare. As a substitute for a complete theory we will,
in this section, provide some heuristic arguments for
the (non-)occurrence of isospectrality which also may give more insight into isospectrality
than detailed
proofs. We think that these arguments could be made rigorous as far as necessary
conditions for isospectrality in  the cases $s=\frac{1}{2}, N=3,4,5$ are involved.(See
the remarks in section 2 on the dimension of isospectral classes and the fibration theorem.)
However,
a detailed proof would require technical issues from the theory of algebraic varieties which
are beyond the scope of this article and, moreover, would appear as superfluous given that
isospectrality in some of these cases has already been proven by case inspection.\\

The heuristic argument goes as follows: Isospectrality will (only) occur if the number
${N \choose 2}$ of bonds between spin pairs exceeds the number $L$ of independent eigenvalues
of $H_0$.
In this case, typically the systems corresponding to an $n={N \choose 2}-L$-dimensional sub-variety
of ${\cal J}$  will possess the same eigenvalues.
The argument may even be
applied if one is not aware of all relations among the eigenvalues which
determine $L$. In such a case of unknown relations one would perhaps over-estimate  $L$
and hence under-estimate the dimension $n$  of the isospectrality classes but, depending
on the case, one could correctly predict the occurrence of isospectrality. \\

When counting the number $L$ of independent eigenvalues one first has to consider the
$(2S+1)$-fold degeneracy dictated by the rotational invariance of $H_0$.
In the simplest example, one has to couple $N=2$ spins $s=\frac{1}{2}$, obtaining one
triplet and one singlet as eigenspaces of $H_0$, symbolically $2\times 2=3+1$.
Thus there are not four, but only two independent eigenvalues of $H_0$.
Similarly, for $N=3$ one has $2\times 2 \times 2=4+2+2$, hence $3$ independent eigenvalues.
In the latter case there are ${3\choose 2}=3$ bonds.  Thus the heuristic
argument does not yet explain isospectrality with $n=1$-dimensional classes. However, it
is easy to find a ``missing relation" among the eigenvalues which reduces the number
of independent eigenvalues to $L=2$: It is just the relation $\Tr H_0 =0$ which yields a
linear relation of the form $4 E_1 + 2 E_2 +2 E_3 =0$ between the eigenvalues $E_\nu$.\\

For $s=\frac{1}{2}$ and arbitrary $N$ it can be shown that there are exactly
${N \choose \left\lfloor N/2 \right\rfloor}$ independent
eigenvalues due to rotational degeneracy and hence,
considering  $\Tr H_0 =0$, $L\le {N \choose \left\lfloor N/2 \right\rfloor} -1$.
In this manner we obtain the results summarized in table 1. \\

\begin{table}\label{T1}
\caption{Occurrence of isospectrality for  $s=\frac{1}{2}$ and arbitrary $N$
based on a heuristic argument on the number of bonds (3rd column)
and the maximal number of independent eigenvalues (2nd column).}
\begin{indented}
\item[]\begin{tabular}{@{}llll}
\br
N & $ {N \choose \left\lfloor N/2 \right\rfloor} -1$ &
${N\choose 2}$ & Isospectrality expected\\
\mr
$2$ & $1$ & $1$ & no\\
$3$ & $2$ & $3$ & yes\\
$4$ & $5$ & $6$ & yes\\
$5$ & $9$ & $10$ & yes\\
$6$ & $19$ &$15$ & no\\
$7$ & $34$ &$21$ & no \\
$\vdots$ & $\vdots$ & $\vdots$ & no \\
\br
\end{tabular}
\end{indented}
\end{table}

\begin{table}\label{T2}
\caption{Non-occurrence of isospectrality for  $s=1$ and small $N$
based on a heuristic argument on the number of bonds (3rd column)
and the maximal number of independent eigenvalues (2nd column).}
\begin{indented}
\item[]\begin{tabular}{@{}llll}
\br
N &$L\le$ &
${N\choose 2}$ & Isospectrality expected\\
\mr
$2$ & $3$ & $1$ & no\\
$3$ & $8$ & $3$ & no\\
$4$ & $24$ & $6$ & no\\
$\vdots$ & $\vdots$ & $\vdots$ & no \\
\br
\end{tabular}
\end{indented}
\end{table}

For large $N$ the the entries in the second column of table 1 grow asymptotically
as $2^N \sqrt{\frac{2}{\pi N}}$, hence almost exponential, whereas ${N \choose 2}$
grows only quadratically. Therefore our heuristic argument will only predict
isospectrality in the cases $s=\frac{1}{2}, N=3,4,5$ but not for larger $N$.
For larger $s>\frac{1}{2}$ the growth of the second column will prevail from the outset,
see table 2, and our argument will not even be applicable for small $N$.
Of course, this does not strictly exclude isospectrality for those cases, but
makes it very unlikely in our opinion.\\

There is another aspect which shows up in table 1 and which we now discuss:
Note that for $N=4$ the difference between the numbers in the 3rd and the 2nd column,
$6-5=1$, would only explain one-dimensional isospectrality classes, whereas we have
encountered two-dimensional classes in section 7. This indicates that the correct $L$
should be $4$, not $5$, and that there is a further ``missing relation" between
the eigenvalues of $H_0$, comparable to $\Tr H_0=0$.
Indeed, as shown in the following paragraph,
there holds a general property of the eigenvalues of arbitrary Heisenberg
Hamiltonians $H_0$ corresponding to the distribution of the eigenvalues among the
quantum numbers $S$.  \\

Let, as above, denote by ${\Bbb P}_S$ the projector onto the eigenspace
of $\bi{S}^2$ with the eigenvalue $S(S+1)$. Then
\begin{equation}\label{61}
\Tr (H_0 {\Bbb P}_S) =
\sum_{x <y}J_{xy} \Tr(\bi{S}_x\cdot\bi{S}_y  {\Bbb P}_S ).
\end{equation}
Since  $\bi{S}^2$ and hence all ${\Bbb P}_S$ commute with arbitrary permutations
of spin sites, the last factor in (\ref{60}) does not depend on $x,y$ and can be
factored out:
\begin{eqnarray}\label{62}
\Tr (H_0 {\Bbb P}_S) &  = &
\left( \sum_{x<y}J_{xy}\right) \Tr(\bi{S}_1\cdot\bi{S}_2 {\Bbb P}_S )\\
 & = & J   \Tr(\bi{S}_1\cdot\bi{S}_2 {\Bbb P}_S ).
\end{eqnarray}
Being independent of the $J_{xy}$ this factor can be calculated for any suitable
$H_0$,  e.~g.~the one with constant $J_{xy}\equiv 1$,

\begin{equation}\label{63}
\tilde{H_0}= \frac{1}{2} \left( \bi{S}^2 -N s(s+1)  \right),
\end{equation}
which yields, after some computation,
\begin{equation}\label{64}
\Tr (H_0 {\Bbb P}_S) =
J \frac{1}{N(N-1)}\left( S(S+1) -N s(s+1)  \right) \Tr {\Bbb P}_S .
\end{equation}
Hence for all $H_0$, the vectors with the components
$\left(\Tr (H_0 {\Bbb P}_S) \right)_{S=S_{min},\ldots,S_{max}}$
are proportional to the constant vector given by the r.~h.~s.~of (\ref{63})
with $J=1$. This gives a number of $S_{max}-S_{min}=\left\lfloor Ns \right\rfloor$
independent linear equations for the eigenvalues of $H_0$. Of course,
$\Tr H_0 = \sum_S \Tr(H_0 {\Bbb P}_S )=0 $ is a consequence of these equations and must not
be counted seperately. For $s=\frac{1}{2}, N=4,5$ we obtain
$\left\lfloor Ns \right\rfloor =2 $ independent equations, which explains the
two-dimensional classes for $N=4$ we found in section 7, and predicts, at least,
two-dimensional classes for the case $N=5$ not yet analyzed in detail. If this case
would show isospectrality classes of dimension $n>2$ one could try to explain this
by invoking more complicated relations derived from the higher moments,
$\Tr (H_0^k {\Bbb P}_S ), k>1$. However, we will not further pursue this question here.

\section{Conclusion}
We summarize our results in table 3. It is in order to add some remarks on the
possibility of determining the coupling constants ${\bi J}$. Our results on the limits
of uniquely determining the values of ${\bi J}$ in the case of isospectrality does not mean that these
values could not be determined otherwise. First, we did not consider
thermodynamical functions which do not come solely from the partition function, such as
correlation functions, etc. Second, we do not adhere to a positivistic attitude
which would in principle deny the physical reality of unmeasurable quantities.
As in other domains of physics, these parameters could also be determined with the aid of
additional assumptions, e.~g.~based on the symmetry of the molecules and
supported by chemical
considerations, which although plausible have not been confirmed directly. So we think
the situation is different from theories with gauge freedom.

\begin{table}\label{T3}
\caption{Our results for the occurence of various types of isospectrality for
different $N,s$. The proven results are indicated as ``yes" or ``no".}
\begin{indented}
\item[]\begin{tabular}{@{}rrlll}
\br
N & s & Plain Isospectrality & Complete Isospectrality & Covariant Isospectrality\\
\mr
$3$ & $\frac{1}{2}$ & yes & yes & yes\\
$3$ & $>\frac{1}{2}$ & unlikely & no & unlikely\\
$4$ & $\frac{1}{2}$ & yes & yes& no\\
$4$ & $>\frac{1}{2}$ & unlikely & no & no\\
$5$ & $\frac{1}{2}$ & likely & likely & no\\
$>5$ & $\frac{1}{2}$ & unlikely & unlikely & no\\
$>5$ & $>\frac{1}{2}$ & unlikely & no & no  \\
\br
\end{tabular}
\end{indented}
\end{table}

\section*{Acknowledgement}
M.~L.~would like to thank members of Fachbereich Physik of Universit\"at
Osnabr\"uck for their warm hospitality during a visit when a part of
this work was performed. We also acknowledge the financial support of
a travel grant awarded jointly by NSF-DAAD. Finally, it is a pleasure
to thank K.~B\"arwinkel, P.~C.~Canfield, V.~G.~Grachev, S.~Jun, D.~Mentrup, J.~Schnack,
and H.~Spindler
for stimulating and helpful discussions. Ames Laboratory is operated for the United
States Department of Energy by Iowa State University under Contract
No. W-7405-Eng-82.

\appendix
\section{Trace formulae}

Expanding the terms occuring in the trace
\begin{equation}\label{71}
\Tr  H_0^n =\Tr \left( \sum_{x<y} J_{xy} \sum_{i=1}^{3} S_x^{(i)}S_y^{(i)}\right)^n
\end{equation}
and using $\Tr (A\otimes B)=(\Tr A)(\Tr B)$ one ends up with terms of the form
\begin{equation}\label{72}
\Tr  (A_1 \ldots A_\ell), A_\nu\in\{S^{(1)},S^{(2)},S^{(3)} \}.
\end{equation}
Here the spin operators without a site index denote operators operating in the single site
Hilbert space ${\Bbb C}^{2s+1}$, not total spin operators.\\

Let $\ell_i (i=1,2,3)$ denote the number of occurences of $S^{(i)}$ in the product $A_1\ldots A_\ell$.
One can easily show that $\Tr  (A_1\ldots A_\ell)$ is non-zero only if all $\ell_i$ are even or all
$\ell_i$ are odd. We give a list of the simplest cases, where the trace is non-zero:
\begin{eqnarray}\label{73}
\ell=0 & : & \Tr ({\Eins})= 2s+1,\\
\ell=2 & : & \Tr  S^{(i)2}=\sum_{m=-s}^s m^2 = \frac{1}{3}s(s+1)(2s+1),\\
\ell=3 & : & \Tr  S^{(1)}S^{(2)}S^{(3)} = - \Tr  S^{(3)}S^{(2)}S^{(1)}=
\frac{i}{6}s(s+1)(2s+1).
\end{eqnarray}
From this we obtain for $n=2$:
\begin{eqnarray}\label{74}
\Tr H_0^2 & = & \sum_{x<y} J_{xy}^2 \sum_{i=1}^3
\Tr \left(S_x^{(i)2}\otimes S_y^{(i)2}\otimes{{\Eins}}_{N-2}\right)\\
     & =  & \left( \sum_{x<y} J_{xy}^2\right) \frac{1}{3} s^2 (s+1)^2 (2s+1)^N.
\end{eqnarray}
For $n=3$ there occur two kinds of non-zero terms:
\begin{equation}\label{75}
\Tr
\left( S_x^{(1)} S_x^{(2)}S_x^{(3)}\otimes S_y^{(1)} S_y^{(2)}S_y^{(3)}\otimes{{\Eins}}_{N-2} \right)
\end{equation}
and those terms obtained by permutations of $\{ 1,2,3\}$, and
\begin{equation}\label{76}
\Tr
\left( S_x^{(i)2}\otimes S_y^{(i)2}\otimes S_z^{(i)2} \otimes{{\Eins}}_{N-3} \right),
\end{equation}
where $x,y,z$ are pairwise distinct and $i=1,2,3$.
Consequently we obtain:
\begin{eqnarray}\label{77}
\Tr H_0^3 & = & T_1 + T_2, \\
T_1 & = & \sum_{x<y} J_{xy}^3 3!(2s+1)^{N-2} \left(  \Tr  S^{(1)} S^{(2)}S^{(3)} \right)^2\\
 & = & - \sum_{x<y} J_{xy}^3 \frac{1}{6} s^2 (s+1)^2 (2s+1)^N,\\
T_2 & = & 3!\left( \sum_{x<y<z} J_{xy} J_{xz}J_{yz} \right) (2s+1)^{N-3}
\sum_{i=1}^3\left(\Tr  S^{(i)2} \right)^3\\
 & = & \left( \sum_{x<y<z} J_{xy} J_{xz}J_{yz} \right) \frac{2}{3}s^3(s+1)^3(2s+1)^N.
\end{eqnarray}

\setcounter{section}{1}

\section{The exceptional cases}

Here we collect some properties of isospectral classes for $N=4,s=1/2$ which belong to
the exceptional case. Although we did not obtain a complete classification of
all isospectral classes these results may be useful for further studies.\\

The subset ${\cal J}\backslash{\cal R}$, which denotes the complement of ${\cal R}$,
is characterized by the vanishing of all $4\times 4$ submatrices of $I'$, hence it will be
also an algebraic variety. After some computations one shows that
${\cal J}\backslash{\cal R}$ is a union
${\cal J}\backslash{\cal R}={\cal C}_1\cup{\cal C}_2\cup{\cal C}_2$
of three simpler varieties given
by the following equations
\begin{equation}  \label{78}
{\cal C}_1:I_1=D_{12}^2+D_{13}^2+D_{14}^2=0,
\end{equation}
\begin{equation}   \label{79}
{\cal C}_2:I_2 = 3 I_3^2, \mbox{    i.~e.~} S_{12}=S_{13}=S_{14},
\end{equation}
\begin{equation}   \label{80}
{\cal C}_3:f=g=0,
\end{equation}
where
\begin{equation}      \label{81}
  f\equiv  (D_{12}^2 -  D_{14}^2 )D_{13}+
        D_{12} D_{14}(S_{14}-S_{12}) =0,
\end{equation}
\begin{equation}      \label{82}
  g\equiv  (D_{13}^2  -  D_{14}^2) D_{12}+
        D_{13} D_{14}(S_{14}-S_{13}) =0.
\end{equation}

In the first case it is clear that an isospectral class lies entirely inside
${\cal C}_1$ or outside ${\cal C}_1$, since $I_1$ is constant on this class.
An analogous remark applies for ${\cal C}_2$ but the case ${\cal C}_3$ is more
complicated.
In the first case with $I_1=0$ the isospectral classes are $0$-dimensional, since the circles
defined by $I_2=i_2$ and $I_3=i_3$ intersect the variety $I_4=-S_{12}S_{13}S_{14}=i_4$
at most at six points.
In the second case with $S_{12}=S_{13}=S_{14}=\sigma$
the isospectral classes are the $1$-dimensional intersections of the sphere $I_1=i_1, i_1>0$
and the variety
$I_4=2 D_{12}D_{13}D_{14}+(D_{12}^2 +D_{13}^2 +D_{14}^2)\sigma -\sigma^3=i_4$.
In the third case we have shown, using the Eliminate-command of MATHEMATICA$^{\circledR}$,
that $f=g=0$ implies an equation of the form
$P(I_1,I_2,I_3,I_4)=0$, namely
\begin{eqnarray} \label{83}
P & = & -108 I_4^2+4 I_3 I_4(9 I_1+18 I_2-5 I_3^2)\nonumber\\
 & & +(I_1+2 I_2-I_3^2)^2(2 I_2+4 I_2-I_3^2)
\end{eqnarray}
Hence those isospectral classes with $P(I_1,I_2,I_3,I_4)\ne 0, I_1\ne 0, I_2 \ne 3 I_3^2$
lie entirely inside ${\cal R}$ and belong to the generic case. We conjecture that also the
converse holds, namely that $P=0$ implies $f=g=0$ but could not prove it.
By calculating the corresponding Groebner bases, it can be shown that $P$
and $f,g$ generate different ideals in the ring of polynomials in 6 variables.
However,
this will not exclude the possibility that the corresponding real
algebraic varieties may be equal.
Hence we cannot exclude further exceptional cases in the realm $P=0$
but $f\ne 0$ or $g\ne 0$.
The isospectral classes in the case $f=g=0$ will be 1-dimensional: The constraint $f=g=0$ allows
one to express the variables $S_{12},S_{13},S_{14}$ in terms of $I_3,D_{12},D_{13},D_{14}$.
The remaining constraints $I_1=i_1,I_4=i_4$ define a family of curves  obtained as
intersections between spheres  and tube-like surfaces.

\end{document}